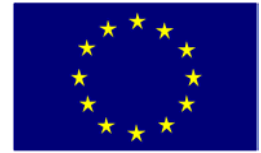

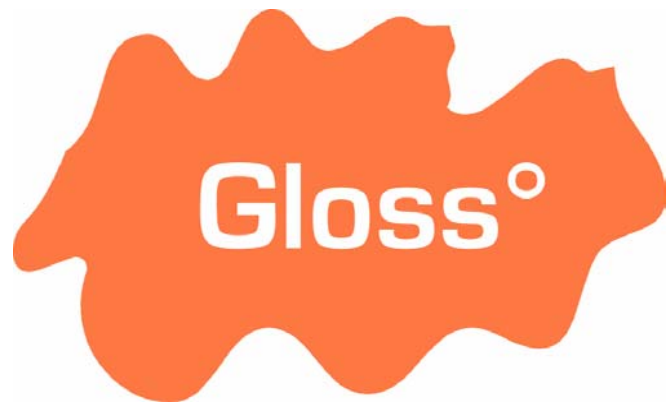

**Global Smart Spaces**

# An Information Flow Architecture for Global Smart Spaces

## D15

### 13 Nov 2003/USTAN/WP6/V1

Alan Dearle
Graham Kirby
Andrew McCarthy
Juan Carlos Dias y Carballo

13/11/03
VERSION 1

| IST Project Number | IST-2000-26070 | Acronym | GLOSS |
|---|---|---|---|
| Full title | Global Smart Spaces | | |
| EU Project officer | Thomas Skordas | | |

| Deliverable | Number | D15 | Name | An Information Flow Architecture for Global Smart Spaces |
|---|---|---|---|---|
| Task | Number | T | Name | n/a |
| Work Package | Number | WP6 | Name | Theories of Mobility |
| Date of delivery | Contractual | PM | Actual | November 2003 |
| Code name | <codename> | | Version | 1.0   draft ☐   final ☑ |
| Nature | Prototype ☐   Report ☑   Specification ☐   Tool ☐   Other: | | | |
| Distribution Type | Public ☑   Restricted ☐   to: <partners> | | | |
| Authors (Partner) | Prof A. Dearle, Dr. G. Kirby, A McCarthy, J Dias y Carballo | | | |
| Contact Person | Prof. A. Dearle | | | |
| | Email | al@dcs.st-and.ac.uk | Phone | +44 1334 463250 | Fax | +44 1334 463278 |
| Abstract (for dissemination) | In this paper we describe an architecture which: Permits the deployment and execution of components in appropriate geographical locations. Provides security mechanisms that prevent misuse of the architecture. Supports a programming model that is familiar to application programmers. Permits installed components to share data. Permits the deployed components to communicate via communication channels. Provides evolution mechanisms permitting the dynamic rearrangement of inter-connection topologies the components that they connect. Supports the specification and deployment of distributed component deployments. | | | |
| Keywords | Information Flow Architecture | | | |

# An Information Flow architecture for Global Smart Spaces

*Introduction*

The Global Smart Spaces project is aimed at supporting the needs of mobile users on a global scale. This requires a large and diverse range of services to be deployed at geographically appropriate locations. Each service is connected to its peers via channels, each carrying some appropriate information. Constantly changing requirements and usage patterns necessitate the ability to introduce new components and change – at runtime – the topology and composition of this environment. That is, both the services running at nodes and the channels by which they are connected must change dynamically.

In order to illustrate the requirements of the information flow architecture, consider Figure 1, which shows a collection of components running on the PDA of a mobile user sending position events via SMS to an SMS server which in turn sends events to a Street Server responsible for some geographical area. The Street Server sends events to a Hearsay service which returns information to the user about items of local interest such as cafes or shops.

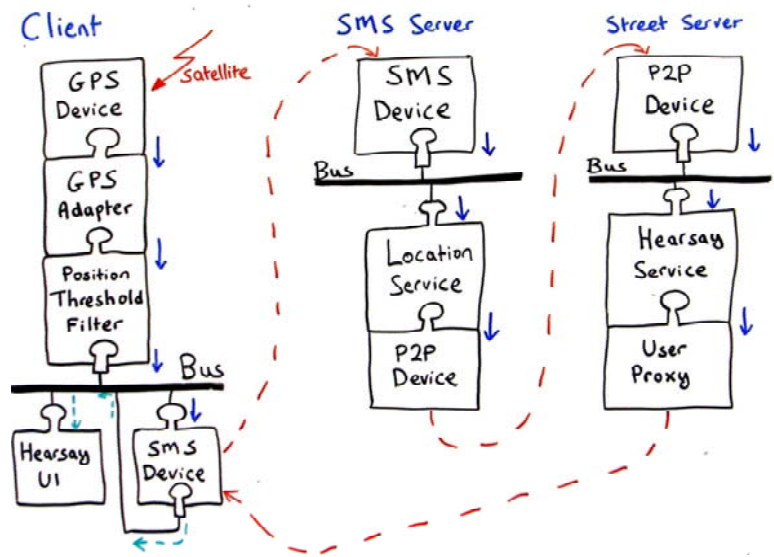

Figure 1: Deployed Components

*Architectural Requirements*

The first requirement is for some kind of architectural description of the components, the hosts which are going to execute the components and the interconnections between the components. In this example, the three hosts hosting computation must be specified: the client, the SMS Server and the Street Server. Next, the set of components running on each host needs to be specified – for example, we need to specify that the Street Server is to run an instance of the P2P Device, The Hearsay Service and the User Proxy. Finally the channels connecting each of the components need to be specified. For example we must specify that the output of the P2P Device component running on the SMS Server needs to be connected to the input of the P2P Device component running on the Street Server.

The second requirement is the ability to enact the architectural description in order to obtain a running deployment consisting of the set of components specified in the architectural description. This requires a number of mechanisms including the ability to both execute and install code on remote servers. The ability to execute and install code on remote machines necessitates a security mechanism to ensure malicious parties cannot execute harmful agents.

To be generally accepted by application programmers, it must be possible to program the architectural components using standard programming languages and appropriate programming models. It should also be possible for architectural components to interface with common off-the-shelf (COTS) components that are already deployed.

In order to permit distributed components to be assembled into appropriate topologies and communicate with each other, the components must exhibit some degree of interface standardisation. In the architecture described here, communication is via asynchronous channels that may be dynamically rebound arbitrary components either by the components themselves or by suitably privileged external parties.

All software architectures are subject to evolutionary pressure – however, we anticipate that architectures supporting mobile users on a global scale will be subject to extreme evolutionary pressures in order to accommodate changes in users' location, activities and interests. This will require the architecture to adapt its topology, caching behaviour, process placement and machine usage.

In order to address these requirements, in this paper we describe an architecture which:
1. Permits the deployment and execution of components in appropriate geographical locations.
2. Provides security mechanisms that prevent misuse of the architecture.
3. Supports a programming model that is familiar to application programmers.
4. Permits installed components to share data.
5. Permits the deployed components to communicate via communication channels.
6. Provides evolution mechanisms permitting the dynamic rearrangement of inter-connection topologies and the components that they connect.
7. Supports the specification and deployment of distributed component deployments.

The system described in this paper is hosted by an enabling infrastructure called CINGAL that supports Computation IN Geographically Appropriate Locations. We describe this system first before describing the GLOSS information flow architecture.

*CINGAL Computational Model*

The CINGAL Computational model is conceptually simple, CINGAL enabled nodes provide: an entry point permitting *bundles* to be *fired*, a content addressable *store*, a name *binder*, an extensible collection of symbolically named *machines* each executing bundles, channel based asynchronous inter-machine communication and a capability system controlling the permissions entities have over stored data, machines and bindings.

A bundle is the only entity that may be executed in CINGAL. Bundles are passive and consist of code, data and a set of bindings naming the data. Each Bundle is uniquely identified by a globally unique identifier (guid) which is implemented via an MD5 key. In practice, Bundles are encoded as XML as shown in Figure 2 below. In the current implementation the code may be written in either Java or Javascript with Java classes being MIME encoded. The code entry point is designated via the *entry* attribute of the *CODE* tag. The data section of a bundle comprises a number of datums each of which has an *id* attribute representing the datum's name within the bundle which must be unique. The collection of *id* names forms a local set of data bindings for the bundle. It is common for Bundles to carry other Bundles as payload. In Figure 2, the Bundle carries another Bundle as a datum called *PAYLOAD*.

```xml
<BUNDLE>
  <CODE entry="uk.ac.stand.dcs.gloss.cingal.Installer" type="java">
      <Class name="uk.ac.stand.dcs.gloss.cingal.Installer">
          DQrK/rq+AAMALQA8CgA
          ... code elided ..
          ACQAAgABACUAAAACACY</Class>
  </CODE>
  <DATA>
    <DATUM id="PAYLOAD">
        <BUNDLE>
            <CODE entry="deployment.deployA.A" type="java">
            <Class name="deployment.deployA.A">
            DQrK/rq+AAMALQAuCgAI
 ... code elided ..
 DwAQAAAAAAACAB0AHgAB</Class>
            </CODE>
            <DATA>
            <DATUM id="LOCAL_Name">Bobs Comp</DATUM>
            </DATA>
        </BUNDLE>
    </DATUM>
    <DATUM id="INSTALL_NAME">ApplicationA</DATUM>
  </DATA>
</BUNDLE>
```

Figure 2: An Example Bundle

Bundles are executed on a remote node by *firing* them. From a remote node this is achieved by sending a bundle to a standard port. Each CINGAL enabled host has a listener that expects clients to communicate with it using the Thin Server Simple Communication Protocol (TSSCP). TSSCP supports a number of operations including the firing of a bundle.

*Machines*

When a bundle is received by the host, provided that the bundle has passed a number of checks described below, the bundle is *fired*, that is, it is executed in a new *machine* as shown in Figure 3. Computation within the fired bundle begins at the entry point specified in the bundle. Bundles may carry out any arbitrary computation that they are encoded to perform including the provision of network services.

Each machine is an isolated protection domain implemented as a separate operating system process. Unlike processes running on traditional operating systems, bundles have a limited interface to their local environment. The repertoire of interactions with the host environment is limited to: interactions with the local store, the manipulation of bindings, the firing of other bundles, and interactions with other machines. Each of these operations is restricted via a capability protection scheme described later.

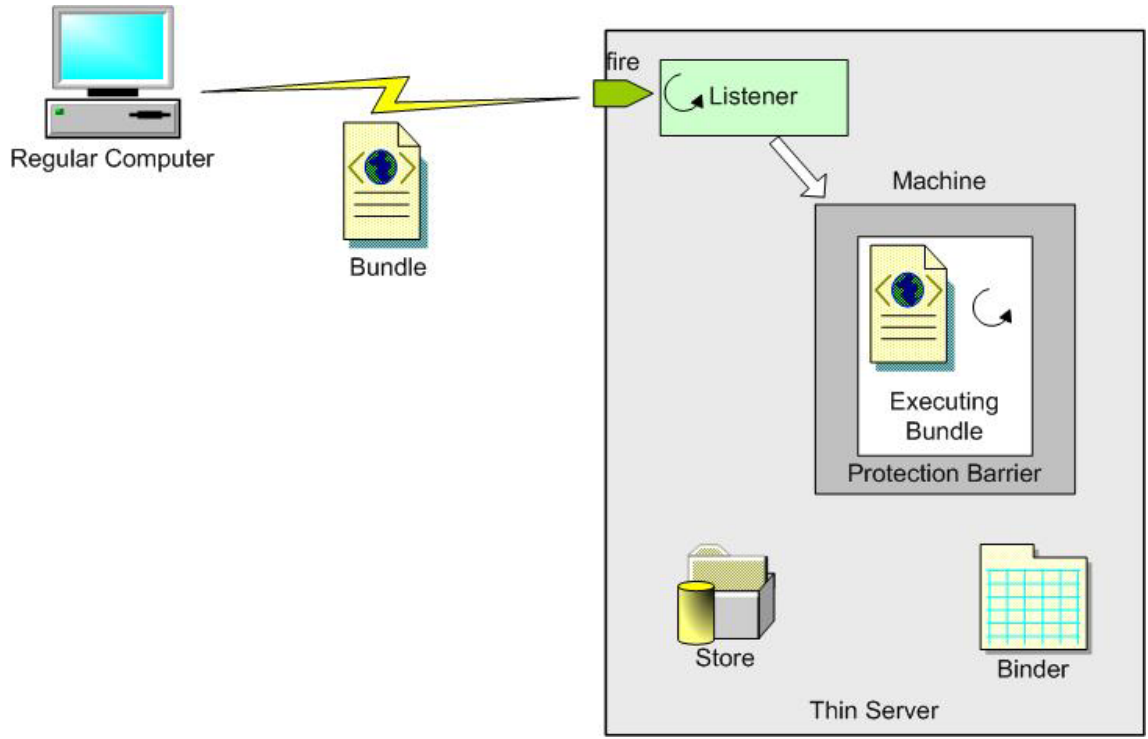

Figure 3: Firing a Bundle

*The Store*

Conceptually the store is a collection of passive data. The *store* component supports the storage of arbitrary bundles. So that a bundle may be retrieved, a *key* in the form of a globally unique identifier (of type *TSGUID*) is returned by the store on its insertion. Stores implement the following interface:

| | |
| ---: | :--- |
| TSBundle | storeGet(TSGUID guid)<br>Retrieves a bundle from the store |
| TSGUID | storePut(TSBundle bundle)<br>Adds a new bundle to the store |
| void | storeRemove(TSGUID guid)<br>Removes a bundle from the store |

Figure 4: The Store Interface

The *storePut* operation inserts a bundle into the store, and returns a key. If that key is later presented via the *storeGet* operation, the original bundle is returned. The *storeGet* operation fails if presented with an unknown key. Stores do not support any update operations. Where the effects of update are required by an application, these may be obtained using *binders* as described later. A desire for simplicity drove the decision to have a store generate the key for a given bit-string, rather than let the key be supplied by the caller.

Figure 5 illustrates the use of the *storePut* and *storeGet* operations to add a bundle and later retrieve it.

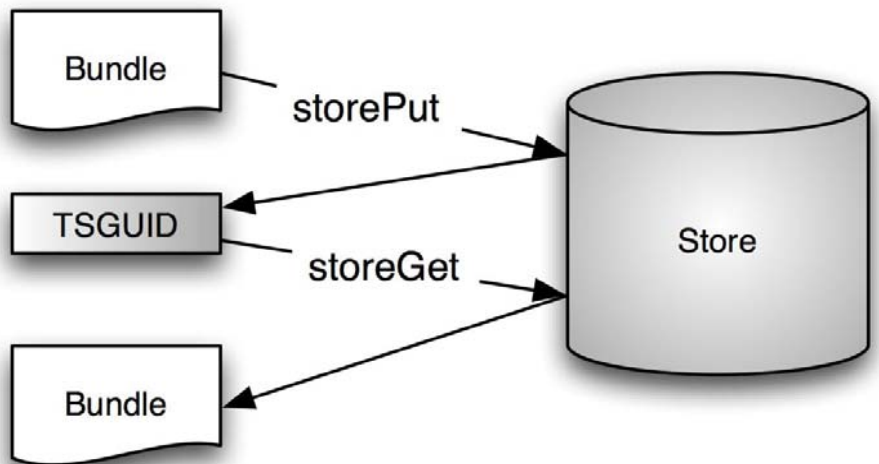

Figure 5: Main Store Operations

*The Store Binder*

The store interface is sufficient to allow information of any kind to be stored and retrieved. For practical use, however, two further abilities are required:

- to support update operations;
- to be able to access stored information through symbolic names as well as arbitrary system-specified keys.

These are provided by the *store binder* (*sbinder*) component, which implements a modifiable many-to-many mapping between symbolic names and keys. A name may be bound to multiple keys, allowing a set to be retrieved in a single operation; a key may be bound to multiple names, giving aliasing. Mappings may be updated so that a given name may refer to various keys over time. The binder provides the following interface:

| TSGUID[ ] | sBinderGet(String symName)<br>Retrieves an entry from the Binder |
|---:|:---|
| void | sBinderPut(String symName, TSGUID guid)<br>Inserts an entry in the Binder |
| void | sBinderPut(String symName, TSGUID guid, String clue)<br>Inserts an entry in the Binder |
| void | sBinderRemove(String symName, TSGUID guid)<br>Removes an entry from the Binder |

Figure 6: Binder Interface

The *sBinderPut* and *BbinderRemove* operations establish and remove a binding between the given name and key respectively. The *sBinderGet* operation returns all the keys currently bound to the given name; this may be an empty set. Figure 7 illustrates the use of the *sBinderPut* and *sBinderGet* operations to bind a name to key binding, and later to retrieve the set of keys currently bound to that name.

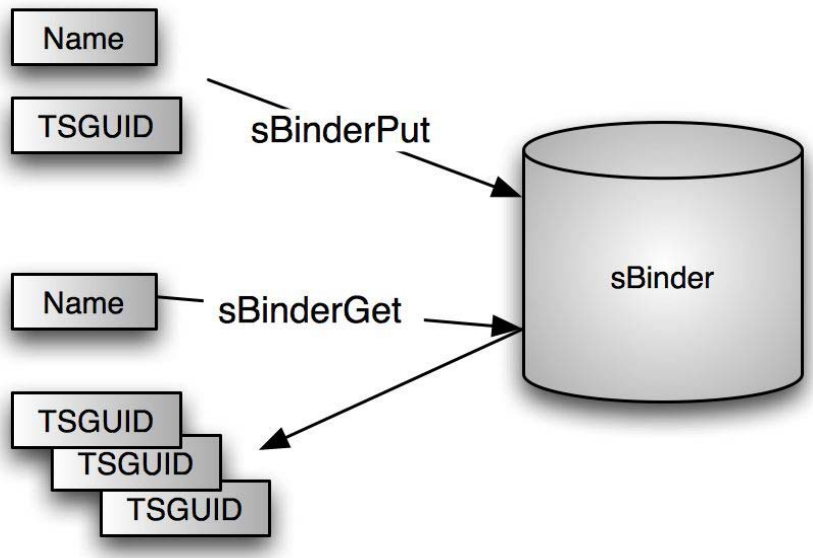

Figure7 Main Binder Operations

*The Process Binder*

Just as the store binder permits data in the store to be symbolically named, another binder, the process binder (pbinder) permits processes (services) to be named. The interface to the pbinder is shown in Figure 8 below. The process binder provides operations for the addition and removal of services via the operations *pBinderPut* and *pBinderRemove*. These operate in a similar fashion to the corresponding store and store binder interface functions. In addition to the specification of a symbolic service name and the GUID of bundle implementing that service, the *pBinderPut* operation takes an extra parameter specifying the maximum number of processes that may be instantiated to deliver the named service. Whenever another process attempts to bind to the named service, a new process will be created to provide that service up to the number specified in *instances* and thereafter the process will be connected to an extant process. Processes lookup services using the inter-process communication mechanisms described below.

| | |
|---:|---|
| void | pBinderPut(String service, TSGUID guid, int instances) <br> Adds a new entry into the store PBinder |
| void | pBinderRemove(String service) <br> Removes an entry from the PBinder |

Figure8 PBinder Interface

A mechanism is also required that permits executing bundles to name themselves as a service. This is achieved using the *setResourceName* method shown in Figure 9.

| | |
|---:|---|
| void | setResourceName(java.lang.String resourceName) <br> Permits a running bundle to name itself as a resource |

Figure9 SetResourceName Interface

*Inter-Process Communication*

CINGAL supports asynchronous message oriented inter-process communication. All communication is via *channels*. In CINGAL, all channels implement the *ITSChannel* interface which supports conventional read and write operations. Whenever a bundle is fired on a machine, a Channel to that bundle is returned to its creator. This Channel may be accessed by the (Thin Server side) fired bundle using the *getDefaultChannel* operation shown in Figure 10 below. In the case of a bundle being pushed from a conventional client, this Channel is returned by the bundle deployment software wrapped up in an object of type *ThinServerClient* as shown below. This channel permits direct communication between a deployed Bundle and its creator, be it local or remote.

| ITSChannel | getDefaultChannel() |
|---|---|
| | The Default user level communications Channel |

Figure10 The get Default Channel Interface

The *ThinServerClient* object is supplied by the CINGAL deployment infrastructure. It provides functionality to interact with a Thin Server. It can connect and send mobile code to a remote Thin Server as well as send requests about resources already present on the local Thin Server. Figure 3 may now be refined as shown in Figure 11 to show the inter-process communication between the deployer and the executing bundle.

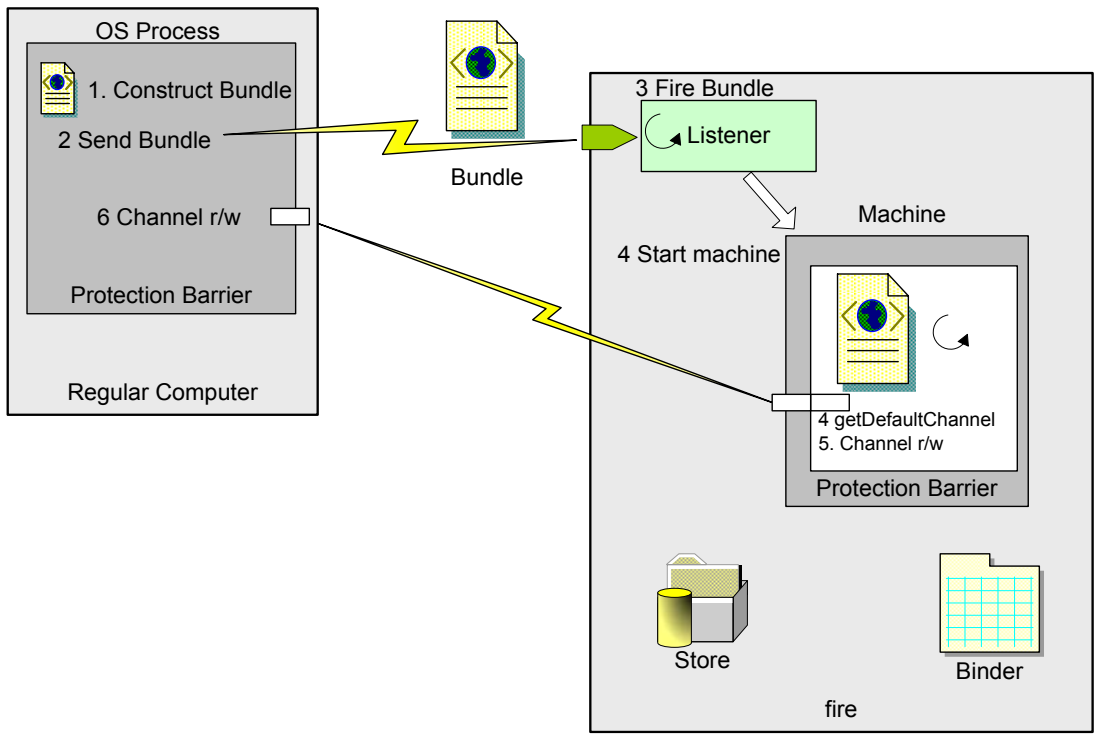

Figure11 Inter-process communication

To fire a bundle, the (almost real) pseudo code shown in Figure 12 is executed.

```
public class OriginatingProcess {
    public void main() {
        // get a bundle from somewhere
        TSBundle bundle = getBundleToFire();
        // send bundle to Thin Server with address ipaddr
        ThinServerClient client = ThinServerClient.send(ipaddr,bundle);
        // Get chanel from ThinServerClient
        TSChannel defaultChannel = client.getResourceChannel();
        // read off the channel
        String messageFromFiredBundle = defaultChannel.getString();
    }
}
```

Figure12 Deployment code

The code executed in the bundle on the Thin Server might look something like that shown in Figure 13 below.

```
public MobileCode implements ITSBundle {
    // Start is the entry point for mobile code
    // This blocks until a connection is made.
    public void Start() {
        ITSChannel defaultChan = machine.getDefaultChannel()
        defaultChan.writeString("HelloWorld");
    }
}
```

Figure13 Deployed Code

The channel established between a bundle and its progenitor is of limited use and is normally only used for diagnostics and the passing of parameters. A running bundle may establish Channels with other bundles using the *resourceConnect* method which has a number of variations shown in Figure 14.

| ThinServerClient | resourceConnect(String resource, String provider) Open a Channel with a resource residing on same TS node. |
| --- | --- |
| ThinServerClient | resourceConnect(String host, String resource, String provider) Open a Channel with a resource residing on another TS node. |
| ITSChannel | resourceConnect(String host, int port) Open a Channel with a resource residing on a conventional node. |

Figure14 Inter-process communication

The first variation of *resourceConnect* is used to obtain a Channel to a resource running on the same Thin Server. The second is used to connect to a process on a remote thin server. In both cases the resource name is used to find a resource which has been registered with the Thin Server either using the *setResourceName* call or has been registered in the pbinder. The last *resourceConnect* call provides a connection to a machine running on a conventional node and functions like a IP socket connection.

*Firing Local Bundles*

Executing Bundles (running in a machine) may instantiate new machines running other bundles using the interfaces shown in Figure 15 below. These permit both bundles from the store and dynamically created bundles to be fired. The latter interface may be used to fire bundles with arbitrary parameters stored in a bundle's data section.

| ThinServerClient | fire(TSGUID guid)<br>Fire a bundle that resides in the node's Store |
|---|---|
| ThinServerClient | fire(TSBundle bundle)<br>Fire a specified bundle |

Figure15 The Local Fire Interface

*Named Channels*

The diagram in Figure 1 shows components arranged in a pipeline with the outputs of one process connected to the inputs of another. The mechanisms describe thus far are sufficient for this purpose. However they lack flexibility in regard to the way in which globally distributed computations may be arranged and evolved. In order to increase this flexibility the CINGAL computational model includes named channels between entities. This idea stems from Milner's pi calculus. Using named channels, individual executing bundles are isolated from the specifics of what components are connected to them. CINGAL provides mechanisms binding, unbinding and rebinding named channels.

When the concept of named channels is introduced, the Bundle execution model may be refined as shown n Figure 16. To recap, the bundle executes within a *machine* presents the bundle with a small number of methods via an interface passed to its start method (the entry point) on initialisation.

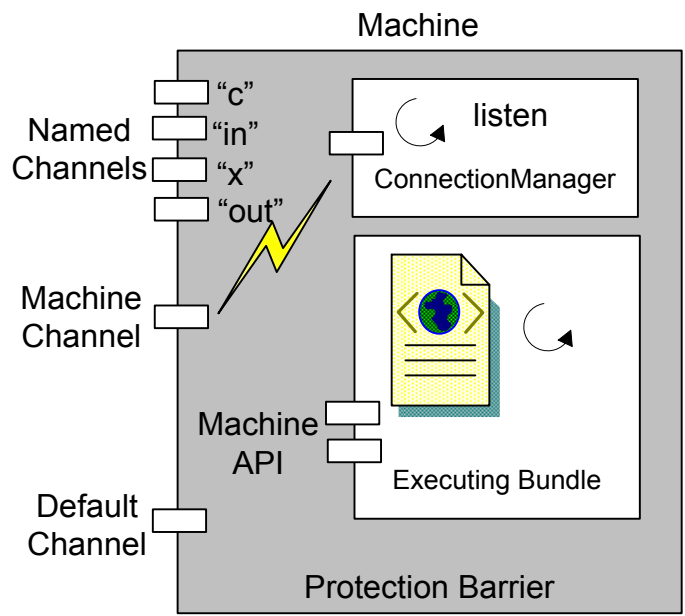

Figure16 The Local Fire Interface

In addition to the operations we have shown this far, the machine interface also permits bundles to obtain bindings to named channels within the machine in which they execute. This interface is called *getAbstractChannel* and takes the name of a named channel as a parameter and returns a channel of type *ITSChannel*.

| | |
|---:|---|
| ITSChannel | getAbstractChannel( String name )<br>Return an abstract Channel |

Figure17 The Abstract Channel interface presented to running Bundles

The channels returned by *getAbstractChannel*() are initially unbound – that is, they are not connected to any other machine instance. Some mechanism needs provided to permit the binding of abstract channels to other channels. Furthermore, the mechanism needs to support the connection of abstract channels by a third party (If this were not so the other channel mechanisms would suffice). To support wiring by third parties, in addition to the default channel, each running machine provides another channel interface which permits interaction with the machine infrastructure rather than the bundle running in the machine. This is channel interface is known as *machine channel* and is also shown in Figure 17. All interactions with the machine are via the TSSCP described earlier.

Within the machine infrastructure a component called the *connection manager* is responsible for the management of named channels. The interface to this component is shown in Figure 18 below. This component is not accessible to bundles executing with the machine.

| | |
|---:|---|
| boolean | connectChannelToName( String host, int port, String Channel name)<br>Connect to another TS Machine |
| ITSChannel | getAbstractChannel( String name )<br>Return an abstract Channel |
| Port_id | listenForConnectionAndBindToChannel( String name )<br>Wait for binding to a remote named channel |

Figure18 Connection Manager interface

The pseudo code for a machine X establishing a channel between a named channel called **in** contained within machine Y and a named channel called **out** on machine Z is as follows. Machine X sends a TSSCP request to the machine Y requesting that a named channel called **in** is established. Machine Y. Machine Y calls the *listenForConnectionAndBindToChannel* method within Machine Y with "in" as a parameter. The connection manager starts a thread listening for an incoming request and returns the port via TSSCP to machine X. Next, machine X sends a TSSCP message to machine Z with parameters "out" and the address of the node hosting Y and the port returned by machine Y.. Machine Z calls the *connectToChannelName* method of Z's connection manager. This connection manager establishes a connection with machine Y and the channel is established. The *getAbstractChannel* method provided by the connection manager is where the channel implementation resides. The code running in the bundle calls this method indirectly when the machine interface method of the same name is called.

*CINGAL Protection Model*

The model described thus far is a perfect virus propagation mechanism. Code may be executed on remote nodes and that code may create new processes, update the store, create name bindings and fire bundles on other thin servers. The CINGAL system implements a two level protection system. The first level restricts the firing of bundles on thin servers; the second restricts what bundles can do when they are running. However, before describing the CINGAL protection model a distinction must be made between the ability to make use of services provided by CINGAL nodes and the ability to deploy and run code on nodes. The use of services running on a CINGAL server such as a Web server is never restricted by the CINGAL security model, it is the firing of bundles on a remote server and the operations that the fired bundles may perform that is subject to security restrictions.

As stated above the first level of restriction is on who is permitted to fire bundles. Clearly a convention Unix or Windows style security model is not appropriate for thin servers which do not have users in the conventional sense. Instead, security is achieved by means of digital signatures and certificates To implement security, each CINGAL node maintains a list of trusted *entities* each associated with a security certificate. This data structure is maintained in a repository called the *Valid Entity Repository* (VER) which presents the interface shown in Figure 19 below.

| String | verPut(byte[] certificate, String type, String subject, Cap Rights) Adds a new entry in the VER |
|---:|---|
| void | verRemove(String entity) Deletes an entity from the VER |
| boolean | verify(Bundle b) Verifies the integrity and authenticity of a Bundle. |

Figure19 Valid Entity Repository interface

Bundles presented for firing from outwith a Thin Server node are required to be signed by a valid entity stored in the valid entity repository. Like many of the data structures maintained by CINGAL nodes, the VER maintains an associative data structure. In the case of the VER this data structure is indexed by the entity id and maps to a tuple including certificates and rights. Operations are provided for adding (*verPut*) and removing (*verRemove*) entities from the repository. Of course these operations are subject to the second protection mechanism which is capability based. An example of a signed bundle is shown in Figure 20.

```
<BUNDLE>
    <AUTHENTICATION
        entity="19730129df7442a5bb5373447eb91509"
        signature="DQowLAIUPFq…BQu1JP5JfO44" />
    <CODE entry="XX">     … </CODE>
    <DATA>     … </DATA>
</BUNDLE>
```

Figure20 A Signed Bundle

The attributes of the AUTHENTICATION tag represent the name of an entity in the VER of the node on which the bundle is being fired and the signature is the signed body of the code payload of the bundle. The Thin Server deployment infrastructure provided for deploying bundles from conventional machines provides programmers with the methods shown in Figure 21 to ease the pain of managing the signing of Bundles.

| Bundle | generateSignature(PrivateKey pKey, Bundle b ) <br> Signs a Bundle with the private key of owner. |
| --- | --- |
| PrivateKey | getPrivateKey(File keystore, String type, String alias, char[] password) <br> Retrieves the private key from a key store. |
| protected boolean | verify(byte[] signed) <br> Verifies the integrity and authenticity of a Bundle. |

Figure21 Operations for managing signing

The signing of bundles and their validation on arrival at thin servers prevents the misuse of Thin Server nodes by unauthorised entities. However it does not prevent a bundle from interfering with other bundles or entities in the binder or store. Ideally, bundle could be totally isolated from each other if they wish giving the illusion that they are the only entities running on a Thin Server node. Conversely, bundles should be able to share resources if required.

To address these needs, the second protection mechanism provided by thin servers is a capability based protection mechanism. In addition to the signatures stored in the VER, Thin Server nodes store segregated capabilities for entities stored in the store, sBinder, pBinder and the VER itself. Whenever a running bundle attempts an operation, the capabilities stored in the VER associated with the entity that invoked the operation are checked. The operation only proceeds if the entity holds sufficient privilege. Further discussion of these mechanisms is beyond the scope of this document.

*Dynamic Deployment using the Deployment Engine*

The CINGAL system provides the infrastructure for deploying arbitrary components in arbitrary geographical components which is a prerequisite for the deployment of Global Smart Spaces. However, it is not sufficient. Some infrastructure needs to be provided to: a) describe global architectures and b) to deploy components from the descriptions. This requirement is addressed by a *description language* and a *deployment engine* and mobile code documents and tools which are described below.

*Deployment Engine*

The Deployment Engine distributes autonomous components which perform a specific computation/function (service). The deployment engine consists of a *parser* that reads a *Deployment Description Document* (DDD) and a *deployer* which transmits bundles which perform the tasks necessary to instantiate the architecture described in the DDD. These tasks typically consist of deploying components, running components and configuring the topology of the deployed application.

*Control Documents*

Central to the deployment process are mobile code tools and XML control documents – *to do lists* and *task reports*. *To do lists* are composed of a set of Tasks which detail actions a tool must attempt to perform upon arrival at a Thin Server. Consequent *task report* documents list the outcomes of each task and any other associated information. When the tool completes its assigned tasks a *task report* is sent back to the *deployment engine*. An example to do list and task report is shown in Figure 22 below.

```xml
<ToDoList>
    <Task guid="urn:gloss:aEcncdeEe" type="INSTALL">
        <datum id="PayloadRef">urn:gloss:a222jdjd2s</datum>
    </Task>
    <Task guid="urn:gloss:aBcbcdebe" type="INSTALL">
        <datum id="PayloadRef">urn:gloss:b333jdjd2s</datum>
    </Task>
</ToDoList>

<TaskReport>
    <TaskOutcome guid="urn:gloss:aEcncdeEe" success="TRUE">
        <!-- TaskOutcomes can have zero, one or many datum
             elements which are bindings and data this
             permits any application specific information
             to be sent back to the Deployment Engine -->
        <datum id="StoreGuid">AECJCJDKSKDLDJSUVDJD</datum>
    </TaskOutcome>
    <TaskOutcome guid="urn:gloss:aBcbcdebe" success="FALSE">
        <datum id="Error">403</datum>
    </TaskOutcome>
</TaskReport>
```

Figure 22 – Example *to do list* and consequential *task report*

Mobile code tools are CINGAL bundles which are configurable by attaching an appropriate to do list to the bundle which encloses the tool. The deployment engine utilises three primary tools: *Installers*, *Runners* and *Wirers*. An example of an installer bundle is shown in Figure 23 below which carries a payload of a bundle containing two Java classes named *MatchingEngine* and *HearsayClient* in addition to the installer code itself which is of of class uk.StAnd....Installer. Note that in addition to the classes, the bundle also carries within its payload a to do list as described above. Note that the payload reference identifies the datum with id="urn:gloss:a222jdjd2s" as the bundle to be installed.

```xml
<BUNDLE>
    <AUTHENTICATION entity="197301m7wWwrPxX9..EySLGU"
       signature="kUdzrv6T..fFNn5Kap" />
    <CODE entry="uk.StAnd....Installer" type="java">
        <Class name="uk.StAnd....Installer"
            <!--  MIME Encoded Class  -->
        </Class>
    </CODE>
    <DATA>
        <DATUM id="urn:gloss:a222jdjd2s">
            <BUNDLE>
                <AUTHENTICATION entity="1973012..91509"
                   Signature="DQowLAIUNs..if1Dn5Kap" />
                <CODE entry="MatchingEngine" type="java">
                    <Class name="MatchingEngine">
                        <!--  MIME Encoded Class  -->
                    </Class>
                    <Class name="MatchingEngine">
                        <!--  MIME Encoded Class  -->
                    </Class>
                    <Class name="HearsayClient">
                        <!--  MIME Encoded Class  -->
                    </Class>
                </CODE>
                <DATA>
                    <DATUM />
                /DATA>
            </BUNDLE>
        </DATUM>
        <DATUM id="ToDoList">
            <ToDoList>
                <Task guid="urn:gloss:aEcncdeEe"
                   type="INSTALL">
                    <datum id="PayloadRef">
                        urn:gloss:a222jdjd2s
                    </datum>
                </Task>
            </ToDoList>
        </DATUM>
    </DATA>
</BUNDLE>
```

Figure 23 An installer

Installer tools install an arbitrary number of bundles into the store of the Thin Server to which they are sent. Runner Tools start the execution of bundles installed in the store of a Thin Server. Wirer Tools are responsible for making concrete connections between pairs of components using the named channel mechanisms described above. Thus components move between three states as they move towards becoming functional components of a deployed architecture:

> **Deployed** – corresponds to the state when a bundle has been installed into the TSStore of a node.
> **Running** – corresponds to the state when a bundle has started computation. Any read/write on named channels will block as they have not been connected.
> **Wired** – corresponds to the state when a bundle has started computation and all abstract channels have been connected to other components.

*Deployment Description Document*

A Deployment Descriptor Document is a static description of a distributed graph of components. An example DDD is shown in Figure XX1. From this example it can be seen that a DDD specifies where to retrieve components (Bundles), the machines available, the mapping of components to machines (a deployment) and the connections between abstract channel pairs. These are specified in the bundles, nodes, deployments and connection sections respectively.

```xml
<DDD name="gloss infrastructure">
   <bundles>
      <bundle name="MatchingEngine"
         code="bundles/MatchingEngine.xml" />
      <bundle name="HearsayCachingServer"
         code="bundles/cachingBundle.xml" />
   </bundles>
   <nodes>
      <node id="als machine" address="129.127.8.34" />
      <node id="andrews machine" address ="129.127.8.23" />
      <node id="grahams machine" address ="129.127.8.35" />
   </nodes>
   <deployments>
      <deployment name ="St_Andrews_Hearsay_Engine"
         bundle="MatchingEngine"
         target="Als Machine" />
      < deployment name ="St_Andrews_Hearsay_Infrastructure"
         bundle="HearsayCachingServer"
         target="andrews machine" />
      < deployment name ="Fife_Hearsay_cache"
         bundle="HearsayCachingServer"
         target="grahams machine" />
   </deployments>
   <connections>
      <connection>
         <source deployment="St_Andrews_Hearsay_Engine"
            channel="OutGoingMatches" />
         <destination
            deployment=
               "St_Andrews_Hearsay_Infrastructure"
            channel="IncomingMatches" />
      </connection>
      <connection>
         <source deployment="Fife_Hearsay_Cache"
            channel="DownstreamCache" />
         <destination
            deployment=
               "St_Andrews_Hearsay_Infrastructure"
            channel="UpstreamCache" />
      </connection>
   </connections>
</DDD>
```

Figure 24 A Deployment Description Document

*The Deployment Process*

The deployment process is as follows, the DDD is input to the Deployment Engine (this process is known as compilation of the DDD). Following compilation, the engine retrieves the specified bundles from a component catalogue and installers are configured (by creating an appropriate to do list). Next the installers are fired (sent to appropriate nodes and executed) to install required components onto Thin Servers throughout the network. One installer is fired per Thin Server. Each installer sends back a report to the deployment engine listing the TSGUIDs of each installed bundle. Figure 25 shows the installation process with the installers running on the set of thin servers specified in the DDD shown in Figure 24.

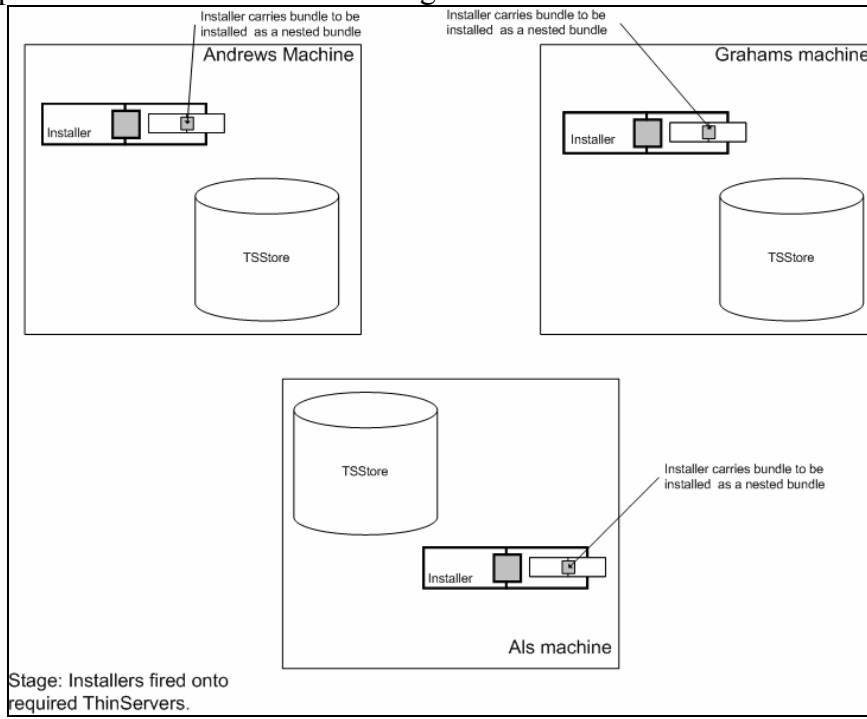

Figure 25 The Installation Process

Upon completion of this phase of the deployment process the required bundles are stored in each Thin Server's store as shown in Figure 26.

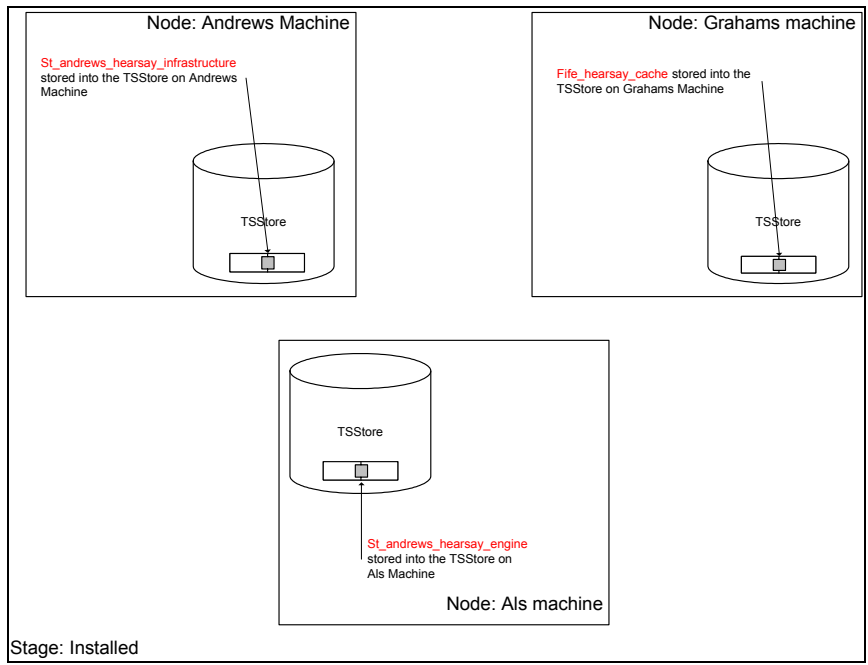

Figure 25 Result of the Installation Process

Each installer tool returns a Task Report to the initiating site containing the store TSGUID for each bundle installed. The example below shows the *task report* returned by the Installer sent to Als Machine.

```xml
<?xml version="1.0" encoding="utf-8" ?>
<TaskReport>
    <TaskOutcome guid="urn:gloss:aEcncdeEe" success="TRUE">
        <datum id="StoreGuid"><!--Store GUID --></datum>
    </TaskOutcome>
</TaskReport>
```
Lists the store GUID the bundle is stored as in the store

Figure 26 Task report sent from Als Machine

Following installation, the deployment engine configures the set of runners by creating an appropriate *to do list*. These runners are fired to start execution of all 'dormant' installed bundles for this deployment. One runner is fired per Thin Server. Figure 27 shows the runner Bundle for Als machine showing the *to do list* instructing the tool to fire a specified bundle from the Store. This bundle is the bundle installed earlier by the installer.

```xml
<BUNDLE>
    <AUTHENTICATION entity="19730129df7442a5bb5373447eb91509"
        signature="DQowLAIUPFqBQu1JP5JfzZO44+CS68HWVRECFGyIm7wWwrPxX9Bu9wu9xZEySLGU">
    </AUTHENTICATION>
    <CODE entry="uk.ac.StAnd.dcs.cingal.deployment.toolkit.runner.Runner" type="java">
        <Class name="uk.ac.StAnd.dcs.cingal.deployment.toolkit.runner.Runner">
            <!-- MIME Encoded Class -->
        </Class>
    </CODE>
    <DATA>
        <DATUM id="ToDoList">
            <ToDoList>
                <Task guid="urn:gloss:32542fdfd444" type="FIRE">
                    <datum id="StoreGuid"><!--Store GUID --></datum>
                </Task>
            </ToDoList>
        </DATUM>
    </DATA>
</BUNDLE>
```

Runner Tool: Classes required for the Runner

ToDoList: Instructs Runner tool to fire specified bundles from the store of the Thin Server it is fired on

Figure 27 The installer bundle sent to Als Machine

Figures 28 and 29 show the runners executing at each Thin Server to fire the appropriate bundles from the Thin Server stores and the result of their execution.

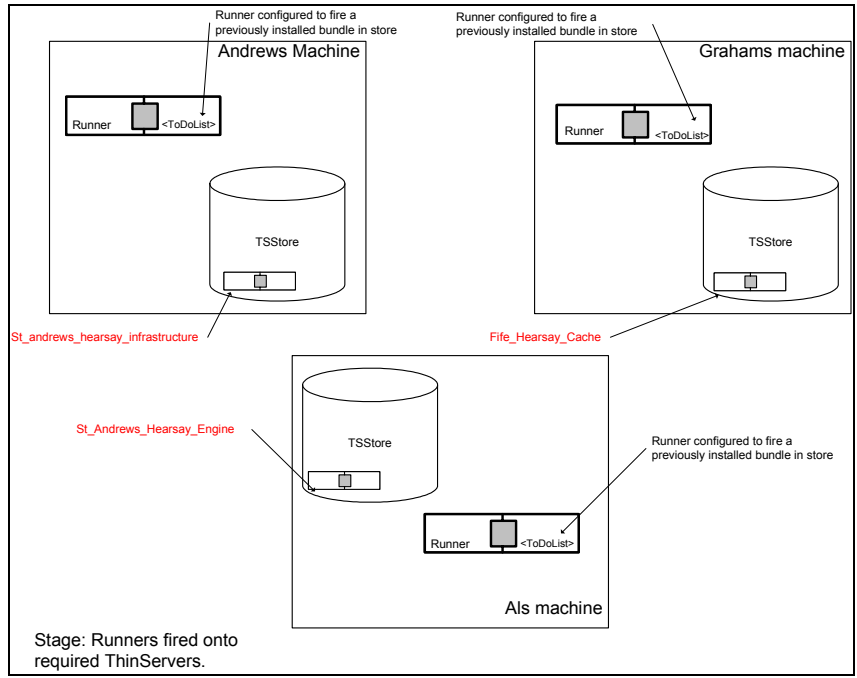

Figure 27 The runner bundles running on the distributed thin servers

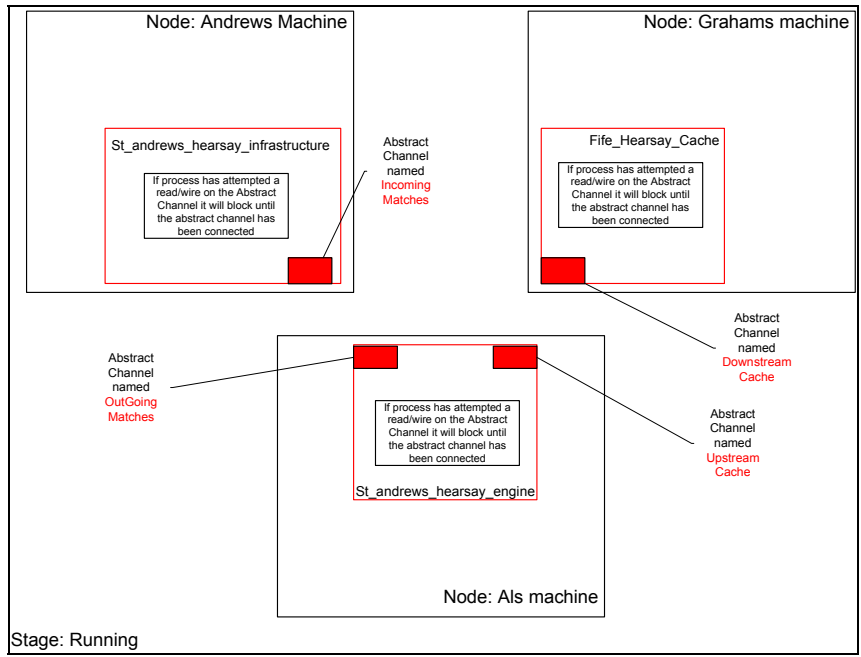

Figure 28 Components running but not connected

As with the installation process, each Runner tool returns a Task Report listing the connector of the enclosing machine for each fired bundle. Figure 29 shows the *task report* returned by the Runner fired on Als Machine.

```xml
<?xml version="1.0" encoding="utf-8" ?>
<TaskReport>
    <TaskOutcome guid="urn:gloss:32542fdfd444" success="TRUE">
        <datum id="Connector">129.127.8.34-3099-4222</datum>
    </TaskOutcome>
</TaskReport>
```

Information about the enclosing machine the bundle was fired on. Allows later communication with the machine or resource.
Format :
<IP Address>-<MachinePort>-<ResourcePort>

Figure 29 Task report sent from Als Machine

The last step in the process is to connect the named channels on each running bundle to assemble the global application topology. To achieve this, the deployment engine first configures wirers by creating appropriate *to do lists* and fires them to connect the named channels in each machine. One wirer is used per connection. The two nodes which hold the channels to be connected are labelled arbitrarily as the primary and secondary nodes. The primary node is where the wiring process will begin, the other end at which the connection is to be created is known as the secondary node.

Each wirer created is provided with configuration data describing
1. The connector for each machine – this contains the IP address of the machine and the machine and resource ports.
2. The name used by executing bundle to reference the channel in both machines (may be different for each machine).

Shown in Figure 30 is the wirer Bundle for Als Machine which is (arbitrarily chosen as) the primary machine for the connection between the named Channels "OutgoingMatches" and "IncomingMatches".

```xml
<BUNDLE>
    <AUTHENTICATION entity="19730129df7442a5bb5373447eb91509"
        signature="DQoewf23rasfgJP5JfzZO44+CS68HWVRECFGyeIm7wWwrPxX9Bu8wu9xZEySLGU">
    </AUTHENTICATION>
    <CODE entry="uk.ac.StAnd.dcs.cingal.deployment.toolkit.wirer.Wirer" type="java">
        <Class name="uk.ac.StAnd.dcs.cingal.deployment.toolkit.wirer.Wirer">
            <!-- MIME Encoded Class -->
        </Class>
    </CODE>
    <DATA>
        <DATUM id="ToDoList">
            <ToDoList>
                <Task guid="urn:gloss:32542xf342444" type="WIRE">
                    <datum id="PrimaryConnector">129.127.8.34-3099-4222</datum>
                    <datum id="SecondaryConnector">129.127.8.23-30112--29000</datum>
                    <datum id="PrimaryAbstractChannel">OutGoingMatches</datum>
                    <datum id="SecondaryAbstractChannel">IncomingMatches</datum>
                </Task>
            </ToDoList>
        </DATUM>
    </DATA>
</BUNDLE>
```

*Wirer Tool: Classes required for Wirer*

*ToDoList: Instructs Wirer Tool to connect the abstract channel "Outgoing Matches" in the Primary Machine to the abstract channel "Incoming Matches" in the secondary machine*

Figure 30 The wirer bundle sent to Als machine

As described above, the Thin Server provides a TSSCP protocol permitting the wirer to communicate with the Connection Manager. The result of the primary phase of the channel establishment process for the working example is shown in Figure 31.

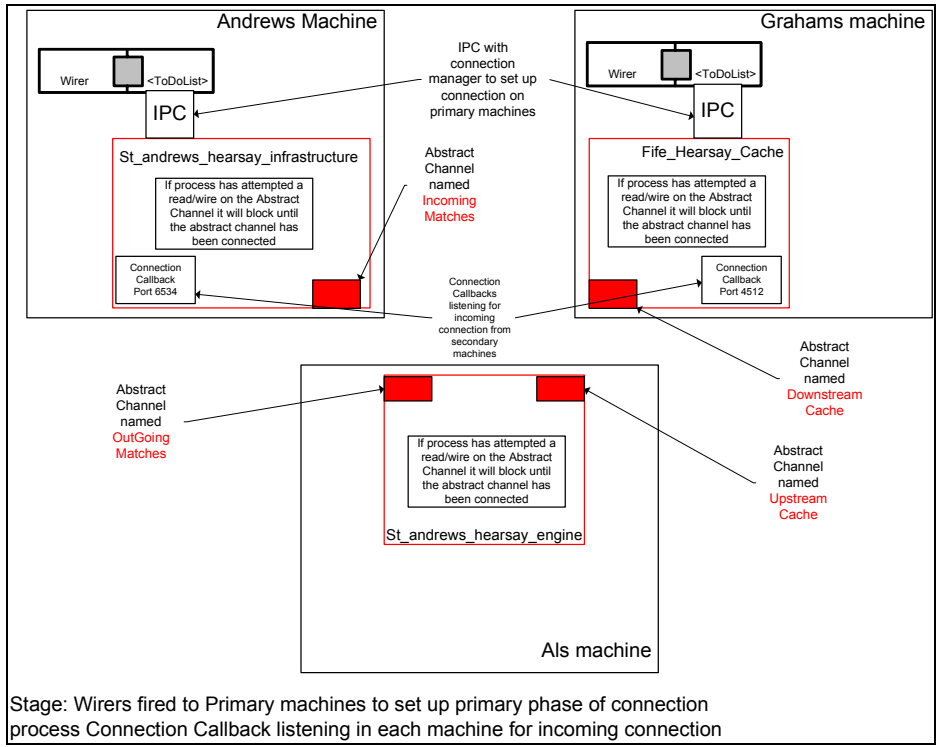

Figure 31 First Phase of Channel establishment

Following completion a listener at the primary node, configures another wirer bundle (its 'offspring') which is sent to the secondary node. The purpose of this wirer is to connect the named Channel on the secondary node to the waiting channel on the primary node. When the offspring wirer arrives at the secondary node, it communicates with the Connection Manager of the machine which requires wiring and instructs it to connect the second named Channel to the listener on the primary node as described above and the connection is established.

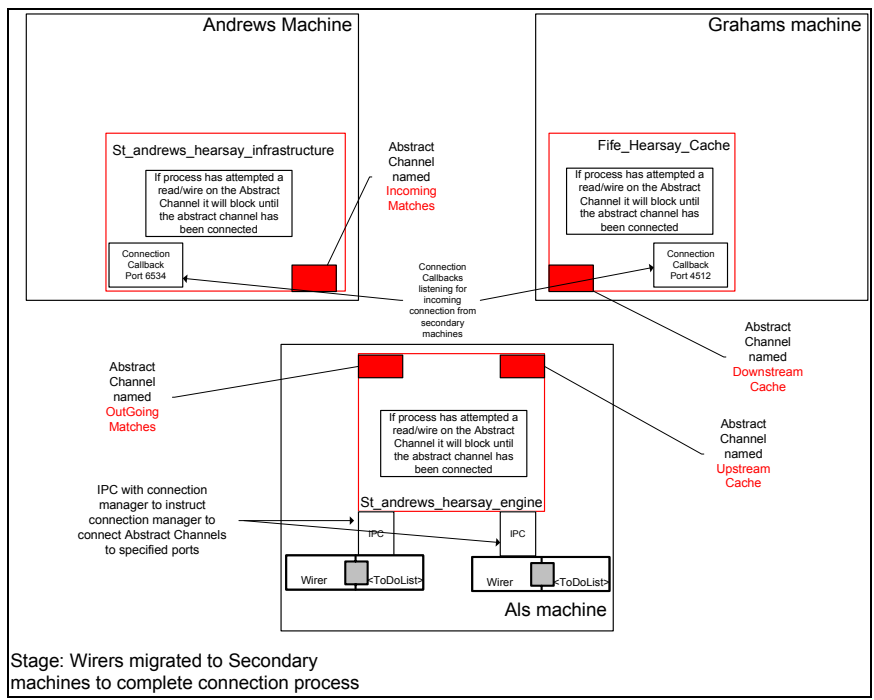

Figure 32 Second Phase of Channel establishment

Once all the wirers have completed (possibly parallel) computation the installation process is complete and all named Channels are connected as shown in Figure 33.

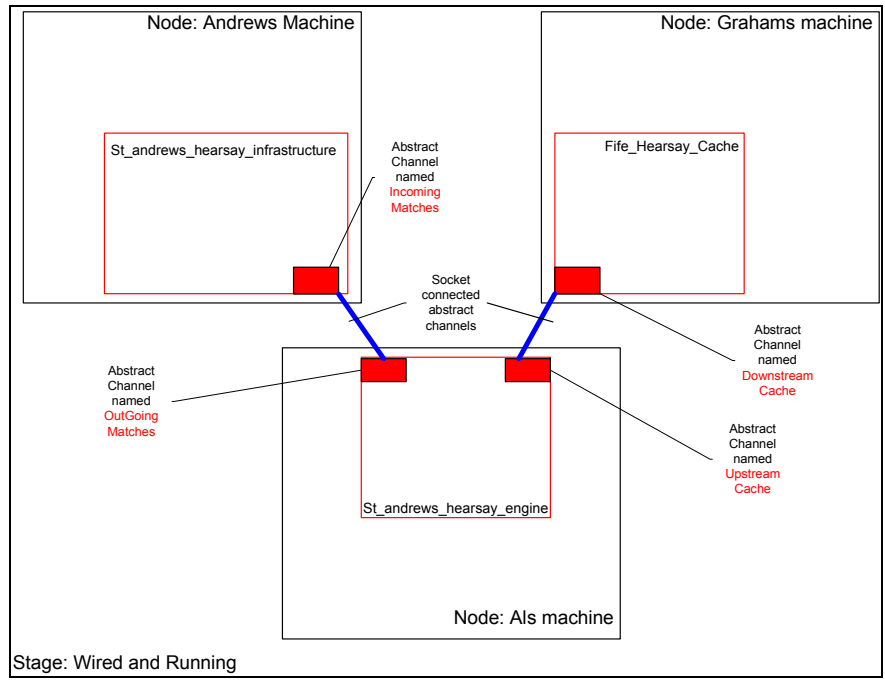

Figure 32 Result of installation

*Conclusion*

At the start of this paper it was claimed that it would describe an architecture which provides the following:

1. Permits the deployment and execution of components in appropriate geographical locations.
2. Provides security mechanisms that prevent misuse of the architecture.
3. Supports a programming model that is familiar to application programmers.
4. Permits installed components to share data.
5. Permits the deployed components to communicate via communication channels.
6. Provides evolution mechanisms permitting the dynamic rearrangement of inter-connection topologies the components that they connect.
7. Supports the specification and deployment of distributed component deployments.

In conclusion these claims are critically re-examined.

*Claim 1*

The CINGAL infrastructure permits bundles to be deployed in arbitrary geographic locations from conventional machines. Bundles may perform arbitrary computation and offer arbitrary network services.

*Claim 2*

The two level security mechanisms provided by CINGAL prevent unauthorised entities from firing bundles on nodes on which they do not have privilege. The ownership model which makes uses of standard cryptographic certificate techniques is well suited to distributed deployment. Tools (not described here) but which operate in a similar manner to the deployment tool are provided for managing entity privileges and updating collections of machines. The capability protection system provided within CINGAL nodes prevents bundles for malicious or unintentional abuse of the Thin Server infrastructure.

*Claim 3*

The programming model is familiar to application programmers. New concepts have been introduced, for example, named channels, but these are not unlike abstractions commonly used by application programmers. Programmers can write bundle code in Javascript or Java and in theory the system could be extended to support programming arbitrary languages.

*Claim 4*

The store and binder provided by Thin Server nodes support content addressed storage which permits code and data to be stored with zero chance of ambiguous retrieval. The binder permits objects to be symbolically named to facilitate the retrieval of components whose content keys are not known. The binder also provides an evolution point supporting update of component mappings.

*Claim 5*

8. Permits the deployed components to communicate via communication channels.

The CINGAL infrastructure supports asynchronous channel based communication. A variety of mechanisms are provided for the establishment of channels including default channels to the progenitors of bundles, standard (socket-based channels) between conventional clients and Thin Server machines and named channels.

*Claim 6*

9. Provides evolution mechanisms permitting the dynamic rearrangement of inter-connection topologies the components that they connect.

A number of novel evolution mechanisms are provided by the architecture. Firstly, the architecture supports the ability to remotely update components. Secondly flexible binding between components is made possible thorough the binder and store interfaces. Most importantly, distributed architectures may be re-arranged by unbinding and reconnecting named channels within running machines running on Thin Server nodes.

*Claim 7*

Distributed Deployment Description documents support the specification of distributed architectures. The deployment engine technology combined with the Thin Server infrastructure permits these distributed deployments to be realised into running instances of component based architectures. The process of deployment from specification through to having a connected collection of running components on distributed nodes is totally automated.

*Future Work*

In the future we propose to expand the system in two primary ways. Firstly we would like to make the specification of distributed components more declarative. To this end we are currently investigating the use of constraint based specification languages. It is our intention to construct higher level specifications and a set of tools to support them and compile these specifications down onto DDD documents. Secondly, we are investigating how evolution can be specified at the DDD level. Since we use DDDs to specify deployments, it seems natural to have high level descriptions of evolution and automatically generate bundles to enact the necessary changes.